# SURFACE STEP EFFECTS ON Si (100) UNDER UNIAXIAL TENSILE STRESS, BY ATOMISTIC CALCULATIONS.

J. Godet, L. Pizzagalli[*], S. Brochard, P. Beauchamp

Laboratoire de Métallurgie Physique, UMR 6630, SP2MI Bld M.& P. Curie 86962, Futuroscope Cedex, France

**Abstract**

This paper reports a study of the step influence at a silicon surface under an uniaxial tensile stress, using an empirical potential. Our aim was to find conditions leading to a nucleation of dislocation from the step. We obtained that no dislocations could be generated with such conditions. This behaviour, different from the one predicted for metals, could be attributed either to the covalent bonding or to the cubic diamond structure.

## 1. Introduction

Dislocation formation at surface defects is a process of particular importance in nanostructured materials submitted to large stresses, such as nanograined systems, or nanolayers in heteroepitaxy for microelectronic devices. The dimension of these materials is typically below few tens of nanometres, which is too small for a Frank-Read mechanism of dislocation multiplication to operate. The dislocations are then likely to have formed at surface or interface defects, such as surface steps. For example, the appearance of dislocations from the cleavage ledge, when silicon is plastically deformed at low temperature, has been recently observed [gally].

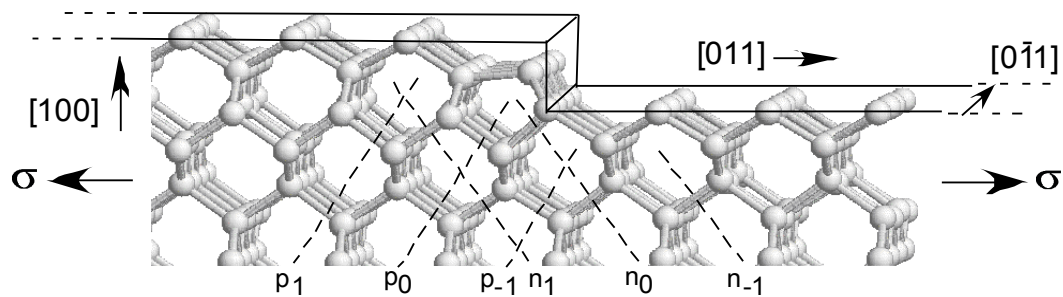

[*] Corresponding author: laurent.pizzagalli@univ-poitiers.fr

Figure 1. Orientation and geometry of the slab for a rebonded $D_B$ step on (100) surface : $\sigma$ is the applied uniaxial stress, $p_i$, $n_i$ point to the <111> planes of the shuffle set.

The very first stages of dislocation formation are still out of the scope of experimental investigations, and atomic scale simulations are expected to bring up useful informations. Several simulations have already been performed, in which for instance the energy of a dislocation is computed as it approaches the free surface [1-3], a dislocation is nucleated from a stressed surface [4-7]. Finite elements calculations have also determined the critical configuration for nucleating the dislocation from the cleavage ledge [Xu]. In the simulations by Brochard *et al.* [8], focussed on metallic systems (Al, Cu), dislocations have formed at the surface step under the simple application of a uniaxial stress. The dislocation is created by localization in a glide plane of a shear whose amplitude and extension increase strongly with the imposed stress. In the present work, a similar study is done, but for silicon, chosen as a model for semiconductors.

**2. Model**

The Si surfaces used for epitaxy or growth have mainly {100} or {111} orientations, easily obtained by cleavage. Here, the {100} surface has been selected owing to its simple atomic configuration compared to the {111} surface. The (100) surface mostly minimizes its energy via a 2x1 reconstruction forming dimers rows [9]. There are two possible step orientations: [011] and [0$\bar{1}$1], along the dense directions, which are the intersection of {111} glide planes and the (100) surface. Depending if one considers a single or double step, two reconstructions are possible. In this study, the most stable step configurations have been chosen.

We modelled the system by a slab with axis along [100], [011] and [0$\bar{1}$1], this last direction corresponding to the surface dimerization (Fig. 1). In these conditions, the step line in the rebonded $S_A$ and $D_B$ configurations is along [0$\bar{1}$1] [10]. Along the surface axis [011] and [0$\bar{1}$1], usual periodic boundary conditions have been applied, which implies the presence of two steps in opposition at the surface. Along the direction perpendicular to the surface, one side of the system (hereafter named bottom) is frozen in order to simulate the silicon bulk.

The system dimensions must be chosen so that the interactions between the two steps or between the free surface and the slab bottom are negligible. Here, different system sizes have been tested, up to 30000 atoms. In the step line direction, the slab exhibit a periodicity of two planes due to the 2x1 reconstruction, but here we used 4 planes for technical reasons. 60 planes have been necessary in the direction normal to the surface, and 64 planes along [011], perpendicular to the steps line.

An uniaxial tensile stress along [011] was applied, i.e. contained in the surface plane and perpendicular to the step line. A homogeneous strain was achieved by fixing atomic positions, including those of the frozen bottom region, as deduced from anisotropic linear elasticity [11], the elastic coefficients being calculated with the atomistic potential. Interatomic interactions were represented by the potential of Stillinger-Weber [12], the large number of atoms, configurations and long relaxation times preventing the use of costly methods such as ab initio techniques. Originally built for describing bulk and liquid phases of Si, the validity of the Stillinger-Weber potential has been proved for small and elastic deformations [13]. Subsequently, in our calculations, we focussed on the effects before the onset of plastic deformations, and we expect that the precursory effects of dislocations nucleation could be detected, as shown by Brochard et al for metals [8].

In order to find the most stable state for each configuration (step geometry, system size, stress value), two methods have been employed. In the first one we performed conjugate gradients minimization for a static relaxation. Then, we also considered molecular dynamics [14]. We let the system evolve at T≠0 during 50 ps in order to enhance the exploration of configurations space. Next, a quench is applied for convergence. For both methods, the convergence is reached when the resulting force on every atoms is less than $10^{-3}$ eV/Å.

Table 1. Evolution of elastic limit for different steps and temperatures. The stresses have been evaluated using linear elasticity.

| Crystal geometry | Bulk | Perfect surface | $S_A$ step | $D_B$ step rebonded | $D_B$ step non rebonded | $D_B$ step rebonded | $D_B$ step rebonded |
|---|---|---|---|---|---|---|---|
| Temperature (K) | 0 | 0 | 0 | 0 | 0 | 300 | 1200 |
| Elastic limit (%) | 32 ± 1 | 28.3 ± 0.5 | 24.7 ± 0.5 | 24.1 ± 0.5 | 22.9 ± 0.5 | 14.3 ± 0.5 | 12.0 ± 0.5 |
| Stress (GPa) | 48 ± 2 | 42.5 ± 0.8 | 37.1 ± 0.8 | 36.2 ± 0.8 | 34.4 ± 0.8 | 21.5 ± 0.8 | 18.0 ± 0.8 |

## 3. Steps and temperature influence on yield stress: plastic effects

The first calculation dealt with a perfect surface at 0 K. Here, the elastic limit corresponded to a strain of 28.3% (Tab. 1), plastic deformations appearing from the bottom of the system. This effect has been explained by the non-physical discontinuity between frozen and relaxed zones. In fact, the applied strain is so large that the system is beyond the linear range of elasticity. For such large deformations a misfit is created between the frozen slab bottom and the relaxed zone. For a bulk system, where there is no frozen part, the calculated elastic limit is larger.

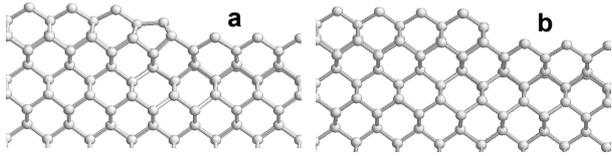
Figure 2. Double height steps: a- rebonded $D_B$ and b- non rebonded $D_B$.

Systems with rebonded $S_A$ and $D_B$ steps on a (100) surface have been tested at 0 K. We have also investigated the effects of a non rebonded $D_B$ for which the step front is more abrupt than the rebonded $D_B$ step (Fig. 2). The Table 1 shows the effects of the step nature on the elastic limit. The presence of steps on the surface decreases the yield stress, compared to the system without step, for all cases. This is a consequence of the symetry breaking and the structure weakening at the step. However, the strain values remain large, which can be explained by the very small concentration of stress at the step, as shown by Poon et al [ ]. It has to be noted that the problem of the discontinuity between frozen and relaxed zones will not have consequences on results for systems with steps because the elastic limit is larger without step than with step. Using different step configurations modifies only slightly the yield stress, about 2%. However, we note that the elastic limit is decreasing when the defect height is increasing, or when the step front is more abrupt.

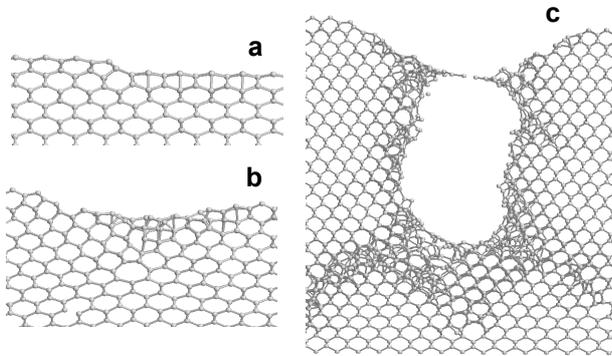
Figure 3. Snapshots showing the evolution of the slab fracture at a rebonded $D_B$ step, at 0 K, and for a deformation of 24.1%.

Additional calculations have been done at non-zero temperatures, in order to introduce more flexibility in the configuration space exploration, and eventually to nucleate a dislocation. Here, only the system with a rebonded $D_B$ step has been studied. This case corresponds precisely to the emergence of a 60° perfect dislocation at the (100) surface, i.e. a step composed of 2 atomic planes. For ambient conditions (300 K), the yield stress is strongly reduced (Tab. 1). The elastic limit decreases again, but slightly, for a temperature of 1200 K. In fact, the higher the temperature, the stronger the thermal vibrations, and the probability to initiate a crack via a bond breaking increases, reducing the elastic limit.

In all cases, the elastic strain increases up to a point (yield) where subsequent deformations (plastic) occur by a depression which forms near the step, in the lower terrace, where there is the maximal stress concentration (Fig. 3a). In this zone, the tension causes bond rupture (Fig. 3b), the system relaxing by appearance of many glide events preferentially along the (111) planes, with the propagation of the crack approximately along the [100] direction, normal to the surface (Fig. 3c). Our calculations have not clearly shown dislocation nucleations from the surface step, even with the introduction of temperature. Still we have shown that the presence of steps on the surface significantly reduces the strength of the system.

## 4. Discussion on elastic behaviour in (111) planes

In the simulation, the silicon sample remains elastic up to a very large elongation, around 25%. There is here a noteworthy difference with the simulation for fcc metals in which a glide occurred at the step around a 8% elongation, preceded by a shear deformation starting from the step and concentrated in the {111} future glide plane [8]. In the absence of surface step, the elastic field in the sample submitted to an uniaxial stress whose direction belongs to the surface plane is homogeneous. The aim here is to investigate the modification of the homogeneous elastic field by the step; one is looking mainly for the occurrence of non linear effects such as {111} shear, characteristic of dislocation pre-nucleation, in the neighbourhood of the step.

The elastic field has been analysed prior to plastic deformation, at 22.3%. The relative displacements between one atom in a (111) plane and its symetric equivalent in the adjacent parallel (111) plane have been calculated in the step region. These displacements includes the effect of the surface relaxation. Only the component along the <211> normal to the step line is considered, since it is the direction of the Burgers vector of an edge dislocation. In the diamond cubic structure, the {111} planes where dislocations can glide are divided into two sets: the widely spaced (1/4<111>) shuffle set planes and the narrowly spaced (1/12<111>) glide set planes [11].

We have considered both a rebonded $D_B$ step and a non rebonded $D_B$ step, even though the latter is not the most stable step configuration. In all cases, the analysis of the relative displacements in the (111) planes of the glide set shows no evidence of shear. In the shuffle set, however, we observed small displacements, having a different sign on both sides of the step, corresponding to a compression on the upper terrace and a tension on the lower terrace (Fig.4 a-b'). This can be explained by the additional planes on the upper terrace, which increase the resistance of the structure.

For the rebonded step, the most important displacements occur in the plane passing through the step edge ($n_0$) (Fig.4 a). This shear would yield to a shortening of the crystal, and is thus in the opposite sense to the dislocation nucleation. This important shear in $n_0$ is not observed for the non rebonded step (Fig.4 b), where the shear of the upper terrace is distributed in several planes $n_{i>0}$. The localisation in a single dense plane for the rebonded step can be directly linked to the rebonding effects.

For planes emerging on the lower terrace with a negative slope (Fig.4 a, b, $n_{i<0}$), the relative displacements are orientated for a dislocation nucleation (elongation), but the shear is not localised in a single plane, which would be a precursor of the nucleation. All these relative displacements are also very small, compared to those observed in metals. Moreover, in the elongated crystal, dislocation nucleation in a plane with a negative slope ($n_i$) would produce an increase of the step height, energetically unfavourable.

In the planes with a positive slope, there is no localisation of shear in a particular plane, even when the shear is in the right sense for dislocation nucleation and step reduction, for both steps (Fig.4 a'b', $p_{i<0}$). Once again, all the relative displacements are very small compared to those calculated in metals. For example, in the $p_{-1}$ plane, the displacement between both planes along <211>, is about 0.02 times the edge component of a perfect 60° dislocation, about one order of magnitude smaller than in metals, and too small for initiating a dislocation.

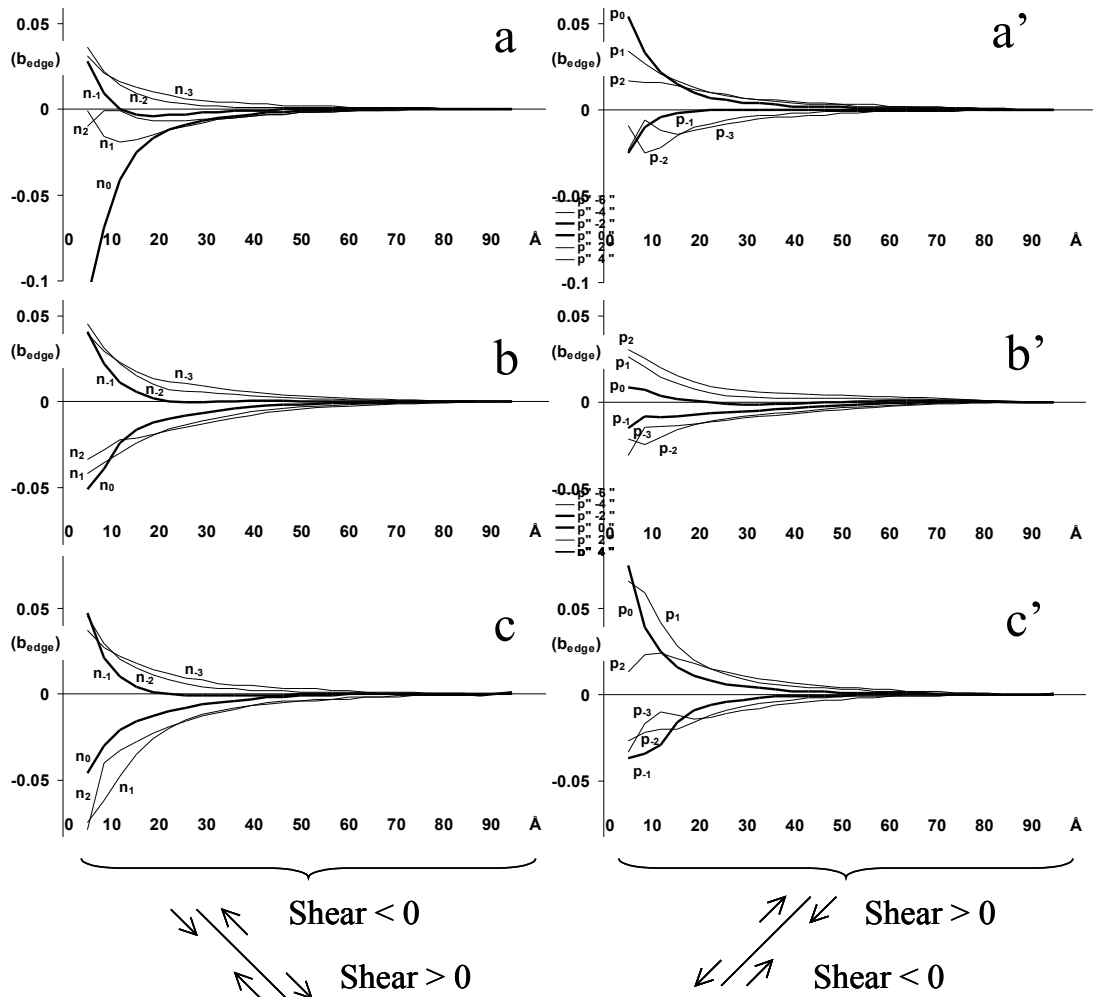

Figure 4. Relative displacements between two adjacent (111) shuffle set planes along <112> direction (unit $b_{edge}$: edge component of perfect 60° dislocation with a/2<110> Burgers vector, a being the lattice parameter) versus the depth in the slab along <112>, a-a' rebonded $D_B$ step, b-b' non rebonded $D_B$ step, c-c' point force. $n_i$ ($p_i$) denote the planes with a negative (positive) slope. (Fig. 1), positive indexes (negative) corresponding to planes from the upper (lower) terrace. Bold curves correspond to the two planes closest to the step edge.

Thus, no precursory shear in the (111) glide plane passing through the step edge (such as $p_{-1}$), eventually helping for the dislocation nucleation, has been obtained for silicon. Using a point force model in which the inhomogeneous elastic field due to the step is produced by a point force applied at the step location on a flat surface, it has been possible to precise the origin of the shear localization for metals: the excess radial compression due to the point force increases, via the Poisson's ratio, the distance between the (111) planes passing through the step edge, resulting in an easier shearing of these planes [8].

It is thus of interest to investigate how the silicon crystal reacts to the application of a point force on the surface. Considering a perfectly flat surface, the point force is located on the surface atoms of the fictitious step line and is directed along [011], i.e. along the direction of the applied uniaxial stress. The force points toward what would be the upper terrace and its amplitude is hσ per length unit, h being the step height and σ the uniaxial stress.

Fig. 4 (c, c') shows that there is no significant difference between the point force model and the step configurations, the relative displacements remaining small and non localised in a single plane. As already observed with the non rebonded step, the point force model confirms that the rebonded effects are negligible, except in the $n_0$ plane. Also, the non appearance of a localised shear precursor of dislocation nucleation can not be due to the step geometry, since it does not appear in the point force case.

There is then a fundamental difference between the behaviours of prototype metal and semiconductor. Since the step geometry is clearly not responsible, two reasons could explain this difference. On the one hand, the silicon cubic diamond structure implies inequivalent (111) planes (shuffle and glide sets), which is not the case for fcc metals. On the other hand, the chemical bonding, covalent for semiconductors and metallic for metals, could be at the heart of the observed differences. However, it remains difficult to estimate the effects of each contribution.

## 5. Conclusion

We have realised an atomistic simulation using an empirical potential to represent a slab of silicon with different surface steps under an uniaxial tensile stress. We have obtained a crack from the lower terrace near the step, for strains about 24% at 0 K. Contrary to metals, it has not been possible to nucleate any dislocation. Before plastic deformations, the upper (lower) terrace is in compression (tension) compared to the bulk. The (111) plane passing through the rebonded $D_B$ step edge shows a concentration of shear opposite to the dislocation nucleation. Using the point force model, we found a behaviour different from the one previously obtained for metals, what could be attributed either to the cubic diamond structure or to the covalent bonding.